\documentclass[nocompress]{spie}  

\usepackage{amsmath,amsfonts,amssymb}
\usepackage{graphicx}
\usepackage[colorlinks=true, allcolors=blue]{hyperref}

\title{Removing grazing incidence reflection with half-bound states and non-Hermitian systems}

\author[a]{Dean A. Patient}
\author[a]{Simon A. R. Horsley}
\affil[a]{University of Exeter, Stocker Road, EX4 4QL, Exeter, Devon}

\authorinfo{Further author information: (Send correspondence to D.A.P.)\\D.A.P.: E-mail: dp348@exeter.ac.uk\\  S.A.R.H.: E-mail: s.horsley@exeter.ac.uk}

\begin{document} 
\maketitle

\begin{abstract}
Grazing incidence waves incident onto a surface will almost always be completely reflected.  Here, we focus on removing reflection at grazing incidence, adopting the factorisation method from quantum mechanics and applying it to the Helmholtz equation that governs a single electromagnetic polarisation.  We show that there are two approaches, the first is to require real dielectric profiles that support a half-bound state at grazing incidence. The second is to allow non-Hermitian dielectric profiles that exhibit PT symmetry, supporting waves with constant intensity throughout the profile. 

\end{abstract}

\keywords{Factorisation Method, Grazing Incidence Reflection, Non-Hermitian}

\section{Introduction}
\label{sec:Intro}  
    Electromagnetic waves incident onto a medium close to grazing incidence ($90^o$) are almost always completely reflected.  This is obvious for a planar interface, where the Fresnel coefficients rapidly increase to unity as incidence approaches grazing.  Although advantageous to improve quantum reflection \cite{Oberst2005}, the behaviour creates problems when reflection is unwanted, such as in the perfectly matched layers of numerical simulations \cite{Chew1996}, and radar absorbers \cite{DeWitt1988}.
    
    There are some notable exceptions to this rule.  For electromagnetic waves, the P\"oschl-Teller potential \cite{Poschl1933} is a permittivity profile that does not reflect transverse electric (TE) radiation at any angle of incidence.  Similarly, complex dielectric profiles obeying the spatial Kramers-Kronig (K-K) relations \cite{Horsley2015} do not reflect any waves of either polarization, and transformation optics (TO) \cite{Pendry2006,Leonhardt2006} generates anisotropic magnetoelectric materials that do not scatter any waves.
    
    It is important that while these examples have theoretically zero reflection at exactly grazing incidence, any real implementation will have imperfections: implemented as a multilayer the P\"oschl--Teller potential has finite steps in permittivity between each layer; a K--K profile must be truncated, with small reflection arising from the truncation; and any TO device will deviate from the ideal $\boldsymbol{\epsilon}=\boldsymbol{\mu}$ condition.  As grazing incidence is a delicate limit, such imperfect implementations have very low reflection in a window \emph{close} to grazing incidence, but then reflect a significant amount at exactly grazing incidence.  The same will be true for the designs given here.
    
    Rather than design a material that is reflectionless for all incident waves, here we focus on the problem of finding material profiles that are reflectionless only for angles close to grazing incidence. In this work, we adopt a factorization method originally used in Quantum Mechanics (QM), finding a family of dielectric profiles that do not reflect Transverse Electric (TE) waves at grazing incidence. 
    
    The factorisation method can be used to derive supersymmetric partner potentials for the Schr\"odinger equation \cite{Cooper1995}: different potentials with the same spectrum, bar the ground state.  As we shall show, the method can also be used to find potentials that support designer bound states.  In optics we can replace the Schr\"odinger equation with the Helmholtz equation, and in that way, the factorisation method can be used to derive optical profiles that support any desired mode.
    
    Here we use the factorization method to find dielectric profiles that support Half-Bound States (HBSs) \cite{Senn1988}, originally discovered as zero energy quantum mechanical states that transmit through a potential, in contrast to the classical intuition that such waves should be completely reflected \cite{Wigner1948}.  The optical equivalent of the HBS is a grazing incidence wave that is transmitted through a dielectric profile without reflection\cite{Patient2021}.  The field of an optical HBS outside of the dielectric is a non-zero constant everywhere. In this way, the mode is `bound' to the dielectric profile, but it cannot be normalised.  The factorisation method can be used to derive an entire class of such non--reflecting profiles.
    
    We also explore complex valued non-Hermitian dielectric profiles, with a distribution of gain and loss. Recently it was shown that such non-Hermitian materials can provide a different route to reflectionless behaviour \cite{Makris2017,Lin2011}. Here we apply the factorisation method to also find complex dielectric profiles, showing  Parity--Time (PT)-symmetric dielectric profiles where there is zero reflection of grazing incidence TE waves.  

\section{Factorisation method}
\label{sec:Factorise}  
\begin{figure} [ht]
    \center
    \includegraphics[width = 0.5\columnwidth]{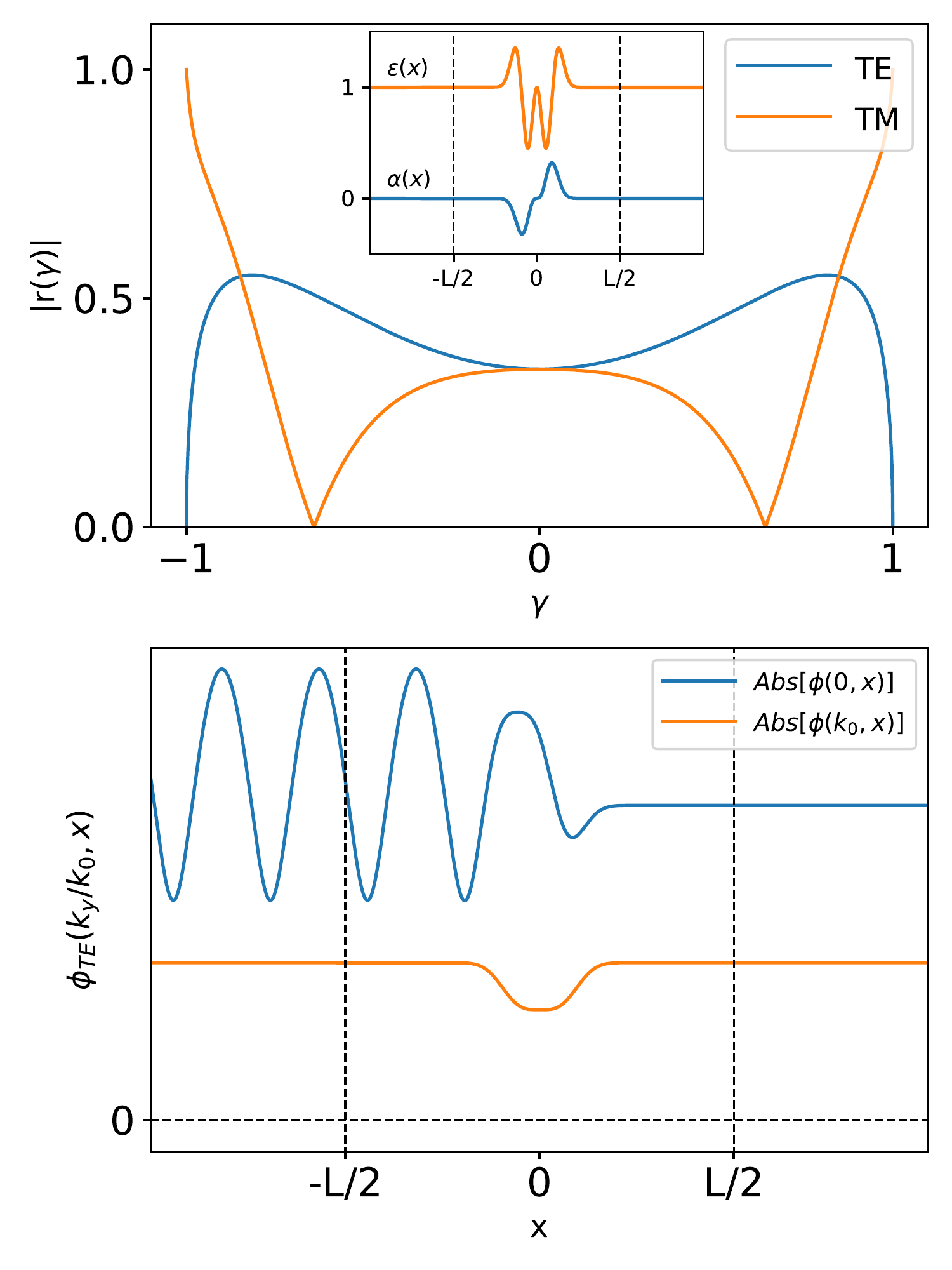}
    \caption{A profile ($\alpha(x) = (x/a_0)^2 \sin(x) \exp(-(x/a_0)^2)$ [$k_0 =1, \lambda = 2\pi/k_0, a_0 = 1, L = 2\lambda$]) is chosen to obtain a dielectric profile (inset) that will not reflect TE waves at grazing incidence (top). The absolute field values at normal and grazing incidence (bottom) demonstrate that at grazing incidence, a HBS is supported.}
    \label{fig:HBS} 
    \end{figure}

    The electric field $E(x,y) = \phi(x) \exp(i k_y y)$ of a TE wave incident at an angle $\theta$ ($k_y = k_0 \sin(\theta)$) onto a 1D graded dielectric profile $\epsilon(x)$ spanning $x \in (\pm L/2)$, flanked on both side by free space, obeys the one dimensional Helmholtz equation
    \begin{equation}
        \label{eq:HH}
        \left( \frac{d^2}{dx^2} + \epsilon(x)k_0^2 - k_y^2 \right)\phi(x) = 0,
    \end{equation}
    for wavenumber $k_0^2 = k_x^2 + k_y^2$.  We assume this equation can be factorised into either of the following two forms 
    \begin{subequations}
        \label{eq:FactorHH}
        \begin{gather}
        \left( -\frac{d}{dx} + k_0 \alpha(x) \right) \left( \frac{d}{dx} + k_0 \alpha(x) \right) \phi(x) = 0, \\
        \left( \frac{d}{dx} + k_0 \alpha(x) \right) \left( -\frac{d}{dx} + k_0 \alpha(x) \right)\phi(x) = 0,
        \end{gather}
    \end{subequations}
    where $\alpha(x)$ is a function that we will choose, and dictates the form of the permittivity function $\epsilon(x)$.  Note that in operator notation, the two equations (\ref{eq:FactorHH}) differ only in the order of the two `raising' and `lowering' operators $\hat{a}=d/dx+k_0\alpha(x)$, and $\hat{a}^{\dagger}=-d/dx+k_0\alpha(x)$.  Expanding out the brackets in Equation~(\ref{eq:FactorHH}) and comparing to Equation~(\ref{eq:HH}), it can be seen that the permittivity profile is related to the function $\alpha(x)$ via
    \begin{equation}
        \label{eq:PermFromAlpha}
        \epsilon(x) = \left(\frac{k_y}{k_0}\right)^2 \pm \frac{1}{k_0}\frac{d \alpha(x)}{dx} - \alpha^2(x).
    \end{equation}
    Given the factorization (\ref{eq:FactorHH}), we can also immediately give the solution to Equation~(\ref{eq:FactorHH}) where either $\hat{a}\phi=0$, or $\hat{a}^{\dagger}\phi=0$.  This is when $(\pm \partial_x + k_0 \alpha(x))\phi(x) = 0$, leading to the solutions
    \begin{equation}
        \label{eq:SUSYSol}
        \phi(x) = \phi(-L/2) \exp\left(\mp k_0 \int_{-L/2}^{x} \alpha(x') dx'\right).
    \end{equation}
    As stated above, a HBS is a mode that is bound to the dielectric profile, but is non-normalisable. This means that, for vanishing reflection at grazing incidence, we require a mode that is a constant with equal amplitude on either side of the dielectric, with some arbitrary mode profile within the confines of the structure.  From Eq. (\ref{eq:SUSYSol}) we can satisfy this condition through requiring that the integral of $\alpha(x)$ is zero over the width of the slab. This would mean that the exponent in (\ref{eq:SUSYSol}) at $\pm L/2$ is zero, and so $\phi(-L/2) = \phi(L/2)$.  An example profile, and the consequent vanishing of grazing incidence reflection is illustrated in Fig. \ref{fig:HBS}. 
    
\section{Non-Hermitian Profiles}
\label{sec:Lossy}  
    We have just shown that the requirement of a HBS solution to the Helmholtz equation can be used to find real valued dielectric profiles with zero grazing incidence reflection.  However the same procedure cannot be applied at an arbitrary angle of incidence, as the wave must be propagating outside of the dielectric, unlike the exponentially decaying solution given in Eq. (\ref{eq:SUSYSol}).   In this case we must make the replacement $\alpha\to{\rm i}\alpha$.  This modifies Equations~(\ref{eq:FactorHH},\ref{eq:PermFromAlpha},\ref{eq:SUSYSol}), with the new form of the factorized Helmholtz equation being
    \begin{subequations}
        \label{eq:FactorHHLoss}
        \begin{gather}
        \left( -\frac{d}{dx} + i k_0 \alpha(x) \right) \left( \frac{d}{dx} + i k_0 \alpha(x) \right) \phi(x) = 0, \\
        \left( \frac{d}{dx} + i k_0 \alpha(x) \right) \left( -\frac{d}{dx} + i k_0 \alpha(x) \right)\phi(x) = 0.
        \end{gather}
    \end{subequations}
    the permittivity taking complex values,
    \begin{equation}
        \label{eq:PermFromAlphaLoss}
        \epsilon(x) = \left(\frac{k_y}{k_0}\right)^2 \pm \frac{i}{k_0}\frac{d \alpha(x)}{dx} + \alpha^2(x)
    \end{equation}
    and the wave propagating through the dielectric without reflection
    \begin{equation}
        \label{eq:SUSYSolLoss}
        \phi(x) = \phi(-L/2) \exp\left(\mp i k_0 \int_{-L/2}^{L/2} \alpha(x') dx'\right).
    \end{equation}

    Interestingly, in the limit of normal incidence, the second solution to Equation~(\ref{eq:PermFromAlphaLoss}) is identical to the non--Hermitian material found by Makris and co--workers \cite{Makris2017}, supporting `constant intensity waves'.  The introduction of the imaginary unit $i$ means that the material is now non--Hermitian, i.e. there is a distribution of loss and gain throughout the profile.  In the particular case of a symmetric function $\alpha(x)$, the permittivity profile (\ref{eq:PermFromAlphaLoss}) is PT--symmetric (see e.g.~\cite{Lin2011}).
    
    Taking the limit of grazing incidence in Eq. (\ref{eq:SUSYSolLoss}) we obtain another different set of materials that do not reflect grazing incidence waves.  We can thus remove grazing incidence reflection in two ways: through either bringing a bound mode from beyond grazing up to the light line (in which case we obtain the lossless material given in Eq. (\ref{eq:PermFromAlpha})), or we can take the angle of the `constant intensity wave' profile in Eq. (\ref{eq:PermFromAlphaLoss}) up to grazing.  
    
\begin{figure} [ht]
    \center
    \includegraphics[width = 0.5\columnwidth]{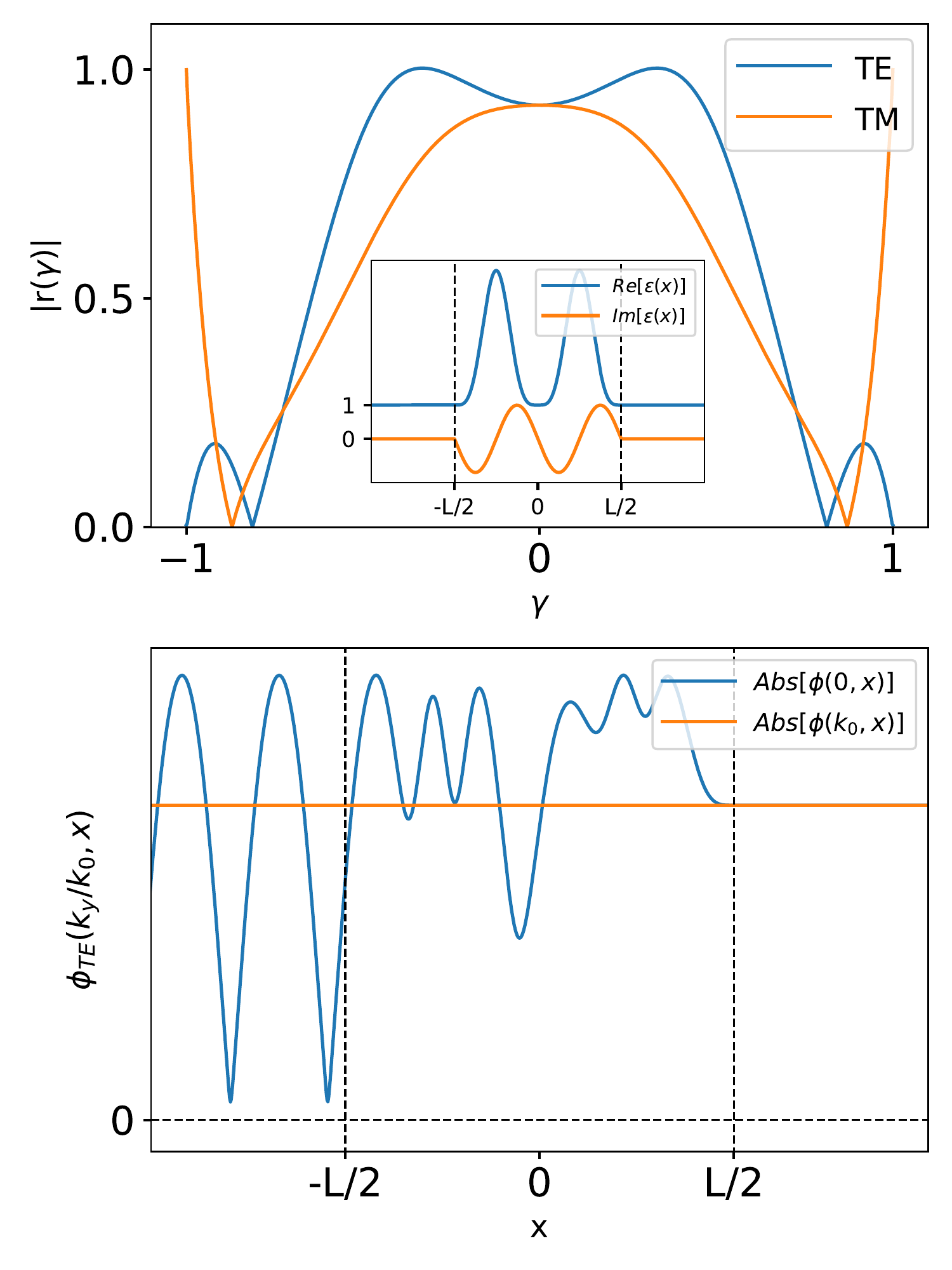}
    \caption{A function $\alpha(x) = \cos(x) - 1$ is used to construct a non--Hermitian permittivity profile $\epsilon(x)$ (inset) that exhibits PT-symmetry, and does not reflect TE waves incident at grazing incidence (top). The resultant field profiles (bottom) show that at grazing incidence, the field intensity $|\phi(x)|$ is constant ($k_0 =1$, $\lambda = 2 \pi/k_0$, $L = 2 \lambda$).}
    \label{fig:PTNoReflect} 
    \end{figure}

    
    Figure~\ref{fig:PTNoReflect} shows a particular example.  To ensure continuity we require that at the boundaries, $\alpha(x)$ should be defined such that it smoothly decays to zero, returning the permittivity (\ref{eq:PermFromAlphaLoss}) to the background value $\epsilon = 1$ at the edges of the profile.   A simple ansatz is a profile of the form $\alpha(x) \sim \cos(x)$.  From this we design a profile that has no reflection of TE waves (Figure~(\ref{fig:PTNoReflect})) is evident at grazing incidence, and the resultant field profiles exhibit a `constant intensity' similar to those shown in \cite{Makris2017}. 
    
\section{Conclusions}
\label{sec:Conclusions}  
    We have shown that through adapting quantum mechanical techniques to the Helmholtz equation we can find a class of both lossless and non--Hermitian materials that do not reflect waves close to grazing incidence.  We showed that a permittivity profile supporting a half--bound state has a transmission resonance at grazing incidence, and hence zero reflection.  Allowing the dielectric profile to be non-Hermitian, we have also shown another different class of complex dielectric profiles that do not reflect grazing incidence TE waves.  We found that this complex version of the HBS is a particular case of the constant intensity waves predicted in \cite{Makris2017}, which has recently been found to be a non--Hermitian instance of a Jackiw--Rebbi mode~\cite{Horsley2019}.

\acknowledgments 
 
The authors would like to acknowledge the Exeter Metamaterials CDT and the EPSRC (EP/L015331/1) for funding and supporting this research. SARH acknowledges financial support from a Royal Society TATA University Research Fellowship (RPG-2016-186).

\copyright Copyright 2022 Society of Photo‑Optical Instrumentation Engineers (SPIE).
According to SPIE Article-Sharing Policies ``Authors may post draft manuscripts on preprint servers such as arXiv. 
If the full citation and Digital Object Identifier (DOI) are known, authors are encouraged to add this information to the preprint record.'' \url{https://www.spiedigitallibrary.org/article-sharing-policies}.

This document represents a draft from Dean A. Patient, Simon A. R. Horsley, "Removing grazing incidence reflection
with half-bound states and non-Hermitian systems," Proc. SPIE 12130,
Metamaterials XIII, 1213003 (24 May 2022)
\url{https://doi.org/10.1117/12.2621826}
Please check out the SPIE paper for a complete list of figures, tables, references and general content.

\bibliography{library} 
\bibliographystyle{spiebib} 

\end{document}